\documentclass[onecolumn,showpacs,plb,superscriptaddress]{revtex4}          
\newcommand{ \be }{\begin{equation}}       
\newcommand{ \ee }{\end{equation}}       
\newcommand{ \bea }{\begin{eqnarray}}       
\newcommand{ \eea }{\end{eqnarray}}

\usepackage{color}

\usepackage{graphicx} 
\usepackage{dcolumn} 
\usepackage{bm} 

\begin{document}          
\title{       
\begin{flushright}  \small \sl version 4,  \today \\  \end{flushright} 
Empirical Constraints on Parton Energy Loss in Nucleus-Nucleus Collisions at RHIC
} 

\author{Gang Wang}
\affiliation{Department of Physics and Astronomy, University of California, Los Angeles, California 90095, USA}
\author{Huan Zhong Huang}
\affiliation{Department of Physics and Astronomy, University of California, Los Angeles, California 90095, USA}
\affiliation{Department of Engineering Physics, Tsinghua University, Beijing, 100084, P.R.China}

\begin{abstract}
We present empirical features of parton energy loss in nucleus-nucleus collisions at RHIC 
through studies of the spectra and nuclear modification factors ($R_{AA}$) for charged hadrons,
neutral pions ($\pi^0$) and non-photonic electrons. The flat distribution of $R_{AA}$ at high transverse momentum ($p_{T}$)
for a given collision centrality is consistent with a scenario where parton energy loss $\Delta p_{T}$
is proportional to $p_{T}$. The centrality dependence of the parton energy loss indicates the absence of
path length dependence in the magnitude of energy loss. The lack of strong path length dependence suggests a 
dynamical picture where the dense partonic medium undergoes rapid expansion and the density of the medium falls rapidly 
in the first a few Fermi interval, which may be much shorter than the full path length. Implications of the empirical 
constraints on the parton energy loss will also be discussed.

\end{abstract}
\pacs{25.75.Ld}
\maketitle

Nucleus-nucleus collisions at RHIC produce a hot and dense medium. Particle production from these collisions
exhibited various distinct physical phenomena depending on transverse momentum ($p_{T}$) scales.
Below $2$ GeV/$c$, hydrodynamic calculations can describe the expansion of the strongly interacting matter
formed in the collisions~\cite{hydrodynamics}.
Between $2$ and $5$ (or $6$) GeV/$c$ particle production exhibits a Constituent Quark Number (CQN) scaling reflecting 
coalescence or recombination hadron formation mechanism with underlying constituent quark degrees of freedom~\cite{coalesence}.
Particle production in the higher $p_T$ region is consistent with features of parton fragmentations,
and may be used to study parton energy loss in the hot and dense medium.
Experimental measurements at RHIC have shown a strong suppression of high $p_T$ particles with respect to 
the expectation of binary nucleon-nucleon collision scaling for charged hadrons~\cite{STAR_charged}, 
for neutral pions~\cite{PHENIX_RAA,PHENIX_RAA_cu} and for non-photonic electrons~\cite{STAR_non_photnic}. 
Early parton energy loss calculations have focused on gluon radiative energy loss mechanism which has been used to explain the high $p_T$ suppression of light hadrons~\cite{Radiative1,Radiative2}. Recent measurements of the strong suppression of
non-photonic electrons from heavy quark decays contradict the expectation of reduced radiative energy loss 
for heavy quarks where a dead-cone effect has been calculated~\cite{kharzeev}. 
It has been proposed that collisional energy loss is important for heavy quarks and cannot be neglected
~\cite{Collisional00,Collisional0,Collisional1,Collisional2}. 
However, a consistent dynamical description of parton energy loss for both light and heavy quarks traversing the hot and dense medium remains illusive. 

It is of great interest to investigate the path length ($L$) and medium density dependence of the energy loss for partons traversing the hot and dense medium created in nucleus-nucleus collisions.
The collisional energy loss per unit length depends on the medium density 
and on the differential cross section weighted by the energy transfer~\cite{Collisional1,Collisional2,Collisional3,AnnualReview,Bass}.
In case of the radiative energy loss, gluons are radiated in
medium-induced multiple scattering processes, 
and the energy loss has been calculated 
to depend, to leading order in $1/E$, on $L$ and 
on the average squared transverse momentum transferred 
to the hard parton per unit length~\cite{Radiative1,Radiative2,AnnualReview,Bass,XN,Baier,Wiedemann}.
Thus both radiative and collisional energy losses are closely related to $L$ and 
the soft parton rapidity density of the medium ($\frac{1}{A_{\perp}}\frac{dN^{g}}{dy}$).
If we assume the linear dependence on $L$ and $\frac{1}{A_{\perp}}\frac{dN^{g}}{dy}$,
both approximately proportional to the cube root of the number of participants ($N_{part}$),
then the energy loss goes with $N_{part}^{2/3}$ as observed for $\pi^0$ in Ref.~\cite{PHENIX_RAA}.
On the other hand, if $\frac{dN^{g}}{dy}$ is approximated by the experimentally measured $\frac{dN}{dy}$,
$A_{\perp}$ by the overlap area of the collision system $S$,
and $L$ by $\sqrt{S}$, then the energy loss goes with $\frac{1}{\sqrt{S}}\frac{dN}{dy}$.
Both $N_{part}$ and $S$ can be estimated with Monte Carlo Glauber calculations~\cite{Glauber}.
In this paper, we extract the effective fractional energy loss for high $p_T$ particles ($5 < p_T < 10$ GeV/$c$) 
using published data from STAR and PHENIX experiments~\cite{STAR_charged,PHENIX_RAA,PHENIX_RAA_cu,STAR_non_photnic},
and study the centrality dependence. The $p_T$ region has been selected so that
we can better focus on the final state medium-induced energy loss since
the initial-state effects, such as nuclear modification of the parton distribution functions
and parton intrinsic $p_T$ broadening, are expected to have small impacts \cite{Eskola,Vitev}.

\begin{figure}[]
  \includegraphics[width=0.450\textwidth]{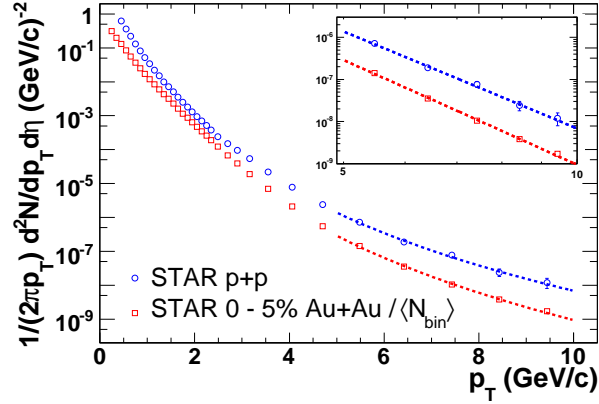}
  \caption{(color online) STAR charged particle spectra in Au+Au ($0 - 5\%$) and p+p collisions at 200 GeV~\cite{STAR_charged}.
		The insert shows the high $p_T$ region in a log-log scale.
		The fitting curves are in the form of the power law function, as discussed in the text.
           }
\label{fig:spectra_demo}
\end{figure}

Nuclear effects on the particle spectra are studied
by comparison with a nucleon-nucleon (NN) reference via
the nuclear modification factor~\cite{TAA}:
\bea
R_{AA}(p_{T}) &=& \frac{d^{2}N^{AA}/dp_{T}d\eta} {T_{AA}d^{2}\sigma^{NN}/dp_{T}d\eta}, 
\label{equ:raa}
\eea
where $T_{AA}=\langle N_{bin} \rangle / \sigma^{NN}_{inel}$ from a Glauber calculation 
accounts for the nuclear collision geometry.
STAR's measurements have shown that for $p_T > 5$ GeV/$c$,
$R_{AA}$ is roughly constant and significantly lower than unity, 
for both charged particles~\cite{STAR_charged} and non-photonic electrons~\cite{STAR_non_photnic}.
The same fact holds for the $\pi^0$ measurements by PHENIX~\cite{PHENIX_RAA,PHENIX_RAA_cu}.

\begin{figure}[]
  \includegraphics[width=0.450\textwidth]{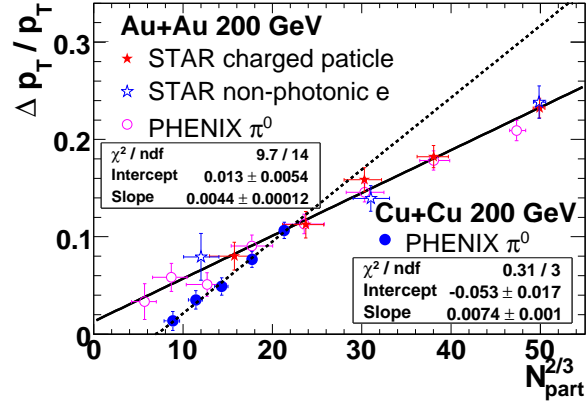}
  \caption{(color online) Fractional energy loss $\Delta p_T/p_T$ (for $p_T > 5$ GeV/$c$) obtained from
Eq.~(\ref{equ:fit}) versus centrality given by $N_{part}^{2/3}$. 
The $R_{AA}$ values are from Refs.~\cite{STAR_charged,PHENIX_RAA,PHENIX_RAA_cu,STAR_non_photnic}. 
The data points for 200 GeV Au+Au (Cu+Cu) collisions are fit with a straight solid (dashed) line.
The $\chi^{2} / ndf = 30 /19$ when we fit all the data points with a single straight line (not shown in the figure).
           }
\label{fig:Energy_loss_Npart}
\end{figure}

In Fig.~\ref{fig:spectra_demo} we show a comparison of STAR's charged particle spectrum
between Au+Au ($0 - 5\%$) and p+p collisions~\cite{STAR_charged},
and fit the data points for $p_T > 5$ GeV/$c$ with the power law function
\be
\frac{1}{2\pi p_T} d^{2}N/dp_{T}d\eta = A(1+p_T/p_0)^{-n}.
\label{equ:power}
\ee
In the insert, fitting functions are straight lines in the log-log plot.
The suppression in the nuclear modification factor can be viewed 
as a horizontal shift effect \cite{PHENIX_RAA} in the spectrum from p+p to Au+Au collisions:
\bea
\frac{d^{2}N^{AA}(p_T)/dp_{T}d\eta} {T_{AA}}&=&\frac{d^{2}\sigma^{NN}(p_T'=p_T+S(p_T))}{dp_{T}'d\eta} \frac{dp_T'}{dp_T}
 = \frac{d^{2}\sigma^{NN}(p_T')}{dp_{T}'d\eta} [1+\frac{dS(p_T)}{dp_T}].
\label{equ:shift}
\eea
With Eq.~(\ref{equ:power}) and Eq.~(\ref{equ:shift}), Eq.~(\ref{equ:raa}) now becomes
\bea
R_{AA}(p_{T}) &=& \frac{(1+p_T'/p_0)^{-n}p_T'}{(1+p_T/p_0)^{-n}p_T}[1+\frac{dS(p_T)}{dp_T}].
\eea
If we assume the fractional $p_T$ shift in the spectrum is a constant, $S(p_T)/p_T=S_0$, then 
\be
R_{AA}(p_{T}) = \frac{(\frac{1}{1+S_0}+p_T/p_0)^{-n}}{(1+p_T/p_0)^{-n}}(1+S_0)^{-n+2}.
\label{equ:fit}
\ee
The parameters $p_0$ and $n$ are obtained and fixed when we fit the spectrum in p+p collisions 
as shown in Fig.~\ref{fig:spectra_demo},
so the $R_{AA}$ of high $p_T$ particles is closely related to $S_0$. In case of small $p_0$ and $S_0$, 
\be
R_{AA} \approx (1+S_0)^{-n+2}
\label{equ:simple}
\ee
is a constant.
In reality, the typical value of $p_0$ is between $0.2$ and $0.3$ GeV$/c$, and $S_0$ is smaller than $0.3$.

\begin{figure}[]
  \includegraphics[width=0.450\textwidth]{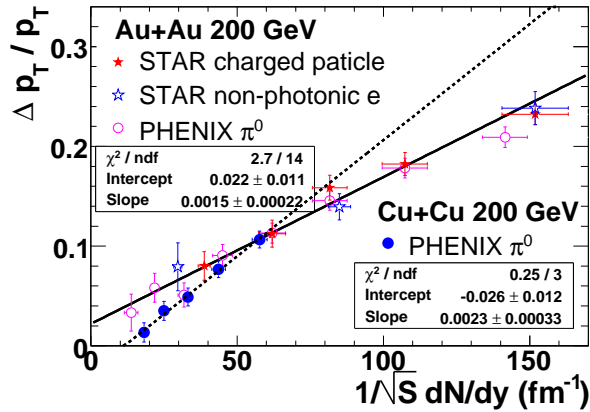}
  \caption{(color online) Fractional energy loss $\Delta p_T/p_T$ (for $p_T > 5$ GeV/$c$) obtained from
Eq.~(\ref{equ:fit}) versus $\frac{1}{\sqrt{S}} \frac{dN}{dy}$. 
The $R_{AA}$ and $\frac{dN}{dy}$ values are from Refs.~\cite{STAR_charged,PHENIX_RAA,PHENIX_RAA_cu,STAR_non_photnic}
and Refs.~\cite{STAR_dndy,PHOBOS_dndy}, respectively.
The fit is in the form of a straight solid (dashed) line for 200 GeV Au+Au (Cu+Cu) collisions.
The $\chi^{2} / ndf = 30 /19$ when we fit all the data points with a single straight line (not shown in the figure).
           } 
\label{fig:Energy_loss_dNdyOverRootS}
\end{figure}

The effective fractional energy loss $\Delta p_T/p_T$ is related to the
fractional shift in the measured spectrum, $S_0$: $\Delta p_T/p_T = S_0/(1+S_0)$.
For $p_T > 5$ GeV/$c$ in 200 GeV Au+Au collisions, we extract the $\Delta p_T/p_T$ 
values by fitting with Eq.~(\ref{equ:fit}) STAR $R_{AA}(p_T)$ results for both charged particles~\cite{STAR_charged} 
and non-photonic electrons~\cite{STAR_non_photnic},
and PHENIX $R_{AA}(p_T)$ result for $\pi^{0}$~\cite{PHENIX_RAA}.
Figure~\ref{fig:Energy_loss_Npart} and \ref{fig:Energy_loss_dNdyOverRootS} show 
the effective fractional energy loss versus centrality given by $N_{part}^{2/3}$ and
$\frac{1}{\sqrt{S}}\frac{dN}{dy}$, respectively.
The values of $\frac{dN}{dy}$ in 200 GeV Au+Au collisions are from Ref.~\cite{STAR_dndy}.
The vertical error bars represent the statistical errors obtained in the fitting procedure.
In both figures, the energy losses of charged particles, $\pi^{0}$ 
and non-photonic electrons, despite different fragmentation functions and decay processes involved for 
these particles, follow the same curve. In parton energy loss scenario, it is generally believed that partons would lose 
energy in the hot and dense medium and the fragmentation process takes place after the partons escape into the vacuum. Assuming that the vacuum fragmentation process is the same in p+p and in A+A collisions, different particles would reflect the same amount of
leading parton energy loss. This scenario is supported empirically by the fact that $\eta$ and $\pi^0$, proton and charged $\pi$ have the same
$R_{AA}$ values at the high $p_T$ region from PHENIX and STAR measurements \cite{PHENIX_eta,STAR_proton}. The semi-leptonic decay kinematics in the heavy quark decays will not change the fact that the transverse momenta of non-photonic electrons are on average proportional to those of heavy quarks.  
Therefore, our extracted $\Delta p_T$/$p_T$ ratios represent the fractional parton energy loss in the medium.

For Au+Au data with various collision centralities the fractional parton energy loss, $\Delta p_T/p_T$, 
increases approximately linearly with $N^{2/3}_{part}$ in Fig.~\ref{fig:Energy_loss_Npart},
and with $\frac{1}{\sqrt{S}}\frac{dN}{dy}$ in Fig.~\ref{fig:Energy_loss_dNdyOverRootS}. 
These $N_{part}$ and $\frac{1}{\sqrt{S}}\frac{dN}{dy}$
dependences are consistent with the scenario of energy loss being determined by path-length-times-density as 
suggested by GLV \cite{GLV} and PQM \cite{PQM}.
However, the accurate Cu+Cu data on $\pi^{0}$ from PHENIX~\cite{PHENIX_RAA_cu} covering the low $N_{part}$ region,
as shown in Fig.~\ref{fig:Energy_loss_Npart} and \ref{fig:Energy_loss_dNdyOverRootS}, systematically deviate from 
the suggested dependence, calling for possible alternative path length and medium density dependence for parton energy loss.
Here, the values of $\frac{dN}{dy}$ in 200 GeV Cu+Cu collisions are converted from 
PHOBOS measurements of $\frac{dN}{d\eta}$~\cite{PHOBOS_dndy},
scaled by a factor of 1.15 from pseudo-rapidity density to rapidity density~\cite{Conversion}.

\begin{figure}[]
  \includegraphics[width=0.450\textwidth]{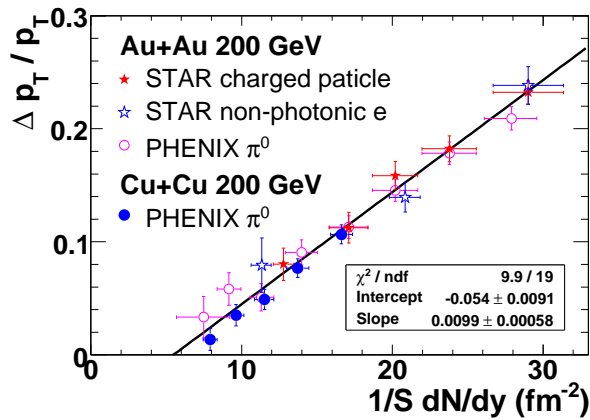}
  \caption{(color online) Fractional energy loss $\Delta p_T/p_T$ (for $p_T > 5$ GeV/$c$) obtained from
Eq.~(\ref{equ:fit}) versus centrality given by $\frac{1}{S} \frac{dN}{dy}$. 
The $R_{AA}$ and $\frac{dN}{dy}$ values are from Refs.~\cite{STAR_charged,PHENIX_RAA,PHENIX_RAA_cu,STAR_non_photnic}
and Refs.~\cite{STAR_dndy,PHOBOS_dndy}, respectively.
The fit is in the form of a straight line for all data points.
When we fit the data points in Au+Au or Cu+Cu separately, the fitting results are consistent with the result shown in the figure.
           }
\label{fig:Energy_loss_dNdy}
\end{figure}

In Fig.~\ref{fig:Energy_loss_dNdy} we show $\Delta p_T/p_T$ as a function
of the particle rapidity density per transverse area ($\frac{1}{S}\frac{dN}{dy}$),
which appears to be a better quantity to describe the dependence of parton energy loss in the hot and dense medium:
the values of fractional parton energy loss, $\Delta p_T/p_T$, from Au+Au and Cu+Cu collisions seem to depend
linearly on $\frac{1}{S}\frac{dN}{dy}$ within statistical errors. 
The fitting line intercepts at a finite $\frac{1}{S}\frac{dN}{dy}$ for zero parton energy loss, possibly indicating that
a minimal medium density and/or transverse dimension of the colliding system is needed in order to have final state energy loss.
We note that the observed $\frac{1}{S}\frac{dN}{dy}$ dependence for parton energy loss implies the absence of or weak path 
length dependence for parton energy loss. The particle rapidity density per transverse area, a quantity likely proportional to
the initial parton density of the collision, may determine the magnitude of the parton energy loss. 
The absence of the path length dependence, if confirmed, 
would contradict calculations of parton energy loss that employ a static average geometry. 

\begin{figure}[]
  \includegraphics[width=0.450\textwidth]{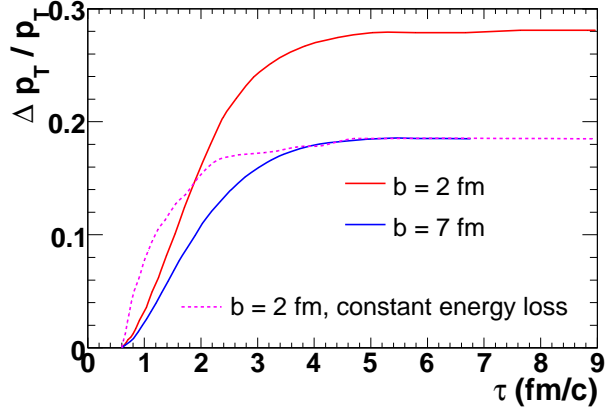}
  \caption{(color online) Fractional energy loss $\Delta p_T/p_T$ (for $4.5 < p_T < 5.5$ GeV/$c$) 
			versus the expansion time of the collision system.
			The $N_{jet}(\tau) /N_{jet}(\tau_{0})$ (or $R_{AA}$) values are from Ref.~\cite{Hydro_EL_time},
			where the dashed line corresponds to the ``constant energy loss" assumption,
and the solid line, the GLV formula \cite{GLV}, as discussed in the text.
           $b = 2$ ($7$) fm corresponds to $0 - 5\%$ ($15 - 25\%$) most central Au+Au collisions, 
	    with $N_{part} \approx 350~(200)$.}
\label{fig:Energy_loss_time}
\end{figure}

Possible physical explanation for the lack of strong path length dependence lies in the fast expansion of the collision system,
leading to a rapidly dropping medium density as a function of time.
We consider the energy loss as a function of the expansion time of the collision system,
in a scenario proposed by Ref.~\cite{Hydro_EL_time} using
a two-component dynamical model (hydro + jet model) \cite{hirano} with a
fully three-dimensional hydrodynamic model \cite{hirano2} for the soft physics
and pQCD jets for the hard physics which are computed via the PYTHIA code \cite{PYTHIA}.
Figure 4 of Ref.~\cite{Hydro_EL_time} shows a hydrodynamic calculation of the jet quenching rate $N_{jet}(\tau) /N_{jet}(\tau_{0})$
for $p_T = 5$ GeV/$c$ jets in Au+Au collisions at 200 GeV,
where $N_{jet}(\tau) /N_{jet}(\tau_{0})$ is equivalent to the $R_{AA}$ of the $p_T = 5$ GeV/$c$ jets
at the expansion time $\tau$.
We apply Eq.~(\ref{equ:simple}) to extract the energy loss information from
the $N_{jet}(\tau) /N_{jet}(\tau_{0})$ (or $R_{AA}$) curves,
and plot the effective fractional energy loss against $\tau$ in Fig.~\ref{fig:Energy_loss_time}.
The dashed line corresponds to the ``constant energy loss" assumption,
that the energy loss per unit length is proportional to the local parton density $\rho(\tau)$ in the medium,
and the solid line, the more sophisticated GLV formula \cite{GLV}.
For both energy loss schemes and both central and mid-central collisions,
a large portion of energy loss occurs within the first a few fm/$c$.
A high $p_T$ parton may traverse the full path length along the trajectory inside a fast expanding medium, but the energy loss
will effectively cease after the first a few fm/$c$ making the total path length irrelevant to the magnitude of the energy loss.
As a result, the effective path lengths are close to a constant for all collision centralities of both Au+Au and Cu+Cu collisions,
and the parton energy losses depend only on the medium density at the early stage of the collision which may be represented by $\frac{1}{S}\frac{dN}{dy}$.
Note that the hydrodynamic calculation is only valid after thermalization of the system.
The actual effective time interval for parton energy loss may be shorter than what the model indicates,
because of non-equilibrium dynamics in the early stage.
Detailed calculations may be needed to address why peripheral Au+Au and Cu+Cu data follow the linear
$\frac{1}{S} \frac{dN}{dy}$ dependence where the participant matter system is not much larger than a few fm in radius.

This physical picture for parton energy loss, if confirmed by more precise data over a broad range of $N_{part}$, would have many implications on theoretical calculations for parton energy loss and on azimuthal angular anisotropy of high $p_T$ particles. Theoretical calculations using static nuclear geometry with average overlapping participant matter will not be able to capture the essence of the limiting time duration for parton energy loss processes. 
Such static calculations, for example, Refs. \cite{Wicks,Eskola2}
concluded that in central A+A collisions the surviving high $p_T$ particles tend to come mostly from surface region of 
the participant matter because partons from the central region would be quenched along the full path length. 
However, our energy loss scenario implies that partons from the central region would suffer energy loss in the first a few fm/$c$ 
and may escape producing high $p_T$ particles in the final state. We argue that the energy loss mechanism is intrinsically a dynamical process and cannot be adequately described by static calculations.  
Theoretical calculations with a three-dimensional hydrodynamic evolution model 
show that an energy loss that grows linear (quadratic) in $L$ 
in a constant medium, as characteristic of collisional (radiative) energy loss, translates into a logarithmic (linear) path length 
dependence in a medium undergoing longitudinal Bjorken expansion~\cite{Bass,Renk1,Renk2,Renk3}. 
In other words, the rapid hydrodynamic expansion weakens the path length dependence of the energy loss.
This theoretical result is consistent with our empirical conclusion that the rapid dynamical 
expansion may account for the path length independence of the energy loss implied by experimental $R_{AA}$ measurements.

Elliptic flow ($v_2$) is defined to be the second-order harmonic of the Fourier expansion of 
particle's $\Delta \phi$ distribution \cite{flow}, 
where $\Delta \phi$ is the particle's azimuthal angle with respect to that of the reaction plane.
The high $p_T$ ($> 5$ GeV/$c$) particle $v_2$, has been considered as arising from path length differences for parton energy loss as a function of $\Delta \phi$.
Our dynamical energy loss scenario would suggest 
that $v_2$ values for high $p_T$ particles would not be as large as predictions based on static energy loss picture. 
The $R_{AA}$ of high $p_T$ $\pi^0$ has been observed to depend on the emission angle 
$\Delta \phi$~\cite{PHENIX_RAA}, an equivalent measurement to that of $v_2$.
The reported angular dependence for $p_T$ $5-8$ GeV/$c$ seems to be weaker than that for $p_T$ $3-5$ GeV/$c$. We cannot draw a firm conclusion on the dynamic scenario for parton energy loss without detailed comparison with theoretical calculations. Significant theoretical and experimental uncertainties arise from large soft particle contributions to the measured $p_T$ range and from the event-plane determination. Both may have biases in the angular distributions.
The event plane is very often estimated with the azimuthal angle distribution
of the detected final-state particles, and ``in-plane" is defined as the direction where most particles come out.
If there are some extra azimuthal correlations (not related to the reaction plane orientation) between
the particles of interest and the ones participating in the event plane determination,
then the measured $v_2$ will deviate from the true value, which is called non-flow effects
including momentum conservation~\cite{momentum conservation}, 
long- and short-range two- and many-particle correlations due to quantum statistics, 
resonances, mini and real jet production, etc~\cite{flow}.
Non-flow effects could easily result in more particles in-plane than out-of-plane,
even if there are no collective motions like elliptic flow or path length dependence of the energy loss
in the high $p_T$ region, especially in the most central and peripheral collisions~\cite{flow200GeV}.
Often to suppress the non-flow effects, the particles used to estimate the event plane are
selected to be one or two units of rapidity away from the particles under study.
But long range correlations may still exist. Further improvement on the event plane determination may use
the side-ward deflected spectator neutrons~\cite{neutron1,neutron2} as carried out in STAR measurements~\cite{v1}.

Another way to study the path length dependence of the energy loss is through the $\Delta \phi$ dependence
of the di-hadron azimuthal correlation after the subtraction of the elliptic flow modulation.
STAR's measurements show that the away-side correlation evolves from single- to double-peak 
with increasing $\Delta \phi$~\cite{Aoqi}, 
where the high $p_T$ trigger particles range from $3$ to $4$ GeV/$c$.
However, one difficulty in this analysis lies in the determination of the elliptic flow background,
which could be larger than the correlation of interest by two orders of magnitude.
Non-flow effects could influence both the measured $v_2$ values and the symmetry of the background correlation,
and flow fluctuations bias the event plane resolution, an important quantity 
to calculate the flow-induced two-particle azimuthal correlations in this analysis~\cite{v2_contribution}.
Further, the Zero Yield at Minimum (ZYAM) approach~\cite{ZYAM} which has often been used 
to normalize the background introduces additional uncertainties.
In addition, the analysis requires large statistics and current results are limited to leading particles 
with $3 < p_T < 4$ GeV/$c$, which is below the $p_T$ domain we considered relevant for parton energy loss study.
Correlation studies triggering on leading particles above $5$ GeV/$c$ are needed to test the physical scenario we have proposed.

In summary, we have extracted the effective fractional energy loss $\Delta p_T/p_T$
for high $p_T$ charged particles, non-photonic electrons and $\pi^0$ in Au+Au collisions
and $\pi^0$ in Cu+Cu collisions at 200 GeV. Empirically the $\Delta p_T/p_T$ is found to be a constant for flat nuclear modification factor 
$R_{AA}$ as a function of $p_T$ from a given collision centrality.
The derived fractional parton energy loss $\Delta p_T/p_T$ seems to depend on particle rapidity density per transverse area for Cu+Cu and Au+Au data over the full range of $N_{part}$, which implies that there is no strong path length dependence for parton energy loss along the trajectory of $N_{part}$ geometry. We argue that the absence of strong path length dependence may be due to the rapid expansion of the participant matter in the early stage so that parton energy loss only takes place within the first a few fm/$c$ duration. Simultaneous studies of nuclear modification factors $R_{AA}$ and elliptic flow $v_2$ for high $p_T$ ($>$ 5 GeV/c) particles would shed more insight on this dynamical scenario for parton energy loss. 

We thank Charles Whitten Jr., Stephen Trentalange and other members of the UCLA Heavy Ion
Physics Group for discussions. This work is supported by a grant from U.S.
Department of Energy, Office of Nuclear Physics.
  

\begin{thebibliography}{99}     
\bibitem{hydrodynamics}
P. F. Kolb and U. Heinz, arXiv:nucl-th/0305084; references therein.
\bibitem{coalesence}
R. J. Fries, V. Greco and P. Sorensen, arXiv:0807.4939 [nucl-th]; references therein.     
\bibitem{STAR_charged}
J. Adams {\it et al.}, \prl {\bf 91}, 172302 (2003).
\bibitem{PHENIX_RAA}
S. S. Adler {\it et al.}, \prc {\bf 76}, 034904 (2007).
\bibitem{PHENIX_RAA_cu}
A. Adare {\it et al.}, arXiv:0801.4555 [nucl-ex].
\bibitem{STAR_non_photnic}
B.I. Abelev {\it et al.}, \prl {\bf 98}, 192301 (2007).
\bibitem{Radiative1}
A. Adil and M. Gyulassy, Phys. Lett. B {\bf 602}, 52 (2004).
\bibitem{Radiative2}
I. Vitev, Phys. Lett. B {\bf 639}, 38 (2006).
\bibitem{kharzeev}
Y.L. Dokshitzer and D.E. Kharzeev, Phys. Lett. B {\bf 519}, 199 (2001).
\bibitem{Collisional00}
H. V. Hees and R. Rapp, \prc {\bf 71} 034907 (2005).
\bibitem{Collisional0}
G. D. Moore and D. Teaney, \prc {\bf 71} 064904 (2005).
\bibitem{Collisional1}
M. G. Mustafa, \prc {\bf 72}, 014905 (2005).
\bibitem{Collisional2}
A. Adil, M. Gyulassy, W. A. Horowitz and S. Wicks,
arXiv:nucl-th/0606010; references therein.
\bibitem{Collisional3}
E. Braaten and M. H. Thoma, \prd {\bf 44}, R2625 (1991).
\bibitem{AnnualReview}
R. Baier, D. Schiff and B. G. Zakharov,
Annu. Rev. Nucl. Part. Sci. {\bf 50}, 37 (2000); references therein.
\bibitem{Bass}
S. A. Bass {\it et al.}, arXiv:0808.0908.
\bibitem{XN}
P. Jacobs and X. N. Wang, Prog. Part. Nucl. Phys. {\bf 54}, 443 (2005); references therein.
\bibitem{Baier}
R. Baier {\it et al.}, Nucl. Phys. B {\bf 483}, 291 (1997).
\bibitem{Wiedemann}
U. A. Wiedemann, Nucl. Phys. B {\bf 588}, 303 (2000).
\bibitem{Glauber}
  B.B. Back {\it et al.},
  \prc {\bf 65}, 031901(R) (2002);
  K. Adcox {\it et al.},
  \prl {\bf 86}, 3500 (2001);
  I.G. Bearden {\it et al.},
  Phys. Lett. B {\bf 523}, 227 (2001);
  J.~Adams {\it et al.},
  arXiv:nucl-ex/0311017.
\bibitem{Eskola}
K.J. Eskola and H. Honkanen, Nucl. Phys. A {\bf 713}, 167 (2003).
\bibitem{Vitev}
I. Vitev and M. Gyulassy, \prl {\bf 89}, 252301 (2002).
\bibitem{TAA}
C. Adler {\it et al.}, \prl {\bf 89}, 202301 (2002).
\bibitem{STAR_dndy}
J. Adams {\it et al.}, \prl {\bf 92}, 112301 (2004).
\bibitem{PHENIX_eta}
S.S. Adler {\it et al.}, \prl {\bf 96}, 202301 (2006).
\bibitem{STAR_proton}
B.I. Abelev {\it et al.}, \prl {\bf 97}, 152301 (2006).
\bibitem{GLV}
I. Vitev, Phys. Lett. B {\bf 639}, 38 (2006).
\bibitem{PQM}
A. Dainese, C. Loizides, and G. Paic, 
Acta Phys. Hung. A {\bf 27}, 245 (2006).
\bibitem{PHOBOS_dndy}
B. Alver {\it et al.}, arXiv:0808.1895 [nucl-ex].
\bibitem{Conversion}
C. Adler {\it et al.}, \prc {\bf 66}, 034904 (2002).
\bibitem{Hydro_EL_time}
T. Hirano and Y. Nara, \prc {\bf 69}, 034908 (2004).
\bibitem{hirano}
T. Hirano and Y. Nara, \prc {\bf 66}, 041901(R) (2002).
\bibitem{hirano2}
T. Hirano, \prc {\bf 65}, 011901 (2002).
\bibitem{PYTHIA}
T. Sjostrand, P. Eden, C. Friberg, L. Lonnblad, G. Miu, S.
Mrenna, and E. Norrbin, Comput. Phys. Commun. {\bf 135}, 238 (2001).
\bibitem{Wicks}
S. Wicks, W. Horowitz, M. Djordjevic and M. Gyulass,
Nuclear Physics A {\bf 784}, 426 (2007).
\bibitem{Eskola2}
K.J. Eskola, H. Honkanen, C.A. Salgado, U.A. Wiedemann,
Nuclear Physics A {\bf 747}, 511 (2005).
\bibitem{Renk1}
T. Renk, \prc {\bf 76}, 064905 (2007).
\bibitem{Renk2}
T. Renk, \prc {\bf 78}, 034904 (2008).
\bibitem{Renk3}
T. Renk, \prc {\bf 78}, 034908 (2008).
\bibitem{flow}
A.M. Poskanzer and S.A. Voloshin, \prc {\bf 58}, 1671 (1998).
\bibitem{momentum conservation}
P. Danielewicz and G. Odyniec, Phys. Lett. B {\bf 157}, 146 (1985).
\bibitem{flow200GeV}
J. Adams {\it et al.}, \prc {\bf 72}, 14904 (2005).
\bibitem{neutron1}
N. Herrmann, J.P. Wessels, and T. Wienold, Annu. Rev. Nucl. Part. Sci. {\bf 49}, 581 (1999).
\bibitem{neutron2}
W. Reisdorf and H.G. Ritter, Annu. Rev. Nucl. Part. Sci. {\bf 47}, 663 (1997).
\bibitem{v1}
J. Adams {\it et al.}, \prc {\bf 73}, 34903 (2006); B.I. Abelev {\it et al.}, arXiv:0807.1518 [nucl-ex].
\bibitem{Aoqi}
A. Feng {\it et al.}, J. Phys. G {\bf 35}, 104082 (2008).
\bibitem{v2_contribution}
J. Bielcikova, S. Esumi, K. Filimonov, S. Voloshin, and J. P. Wurm,
\prc {\bf 69}, 021901 (2004).
\bibitem{ZYAM}
J. Adams {\it et al.}, \prl {\bf 95}, 152301 (2005).
\end{thebibliography}
\end{document}